# Towards the manipulation of topological states of matter:

# A perspective from electron transport

Cheng Zhang[1,2], Hai-Zhou Lu[3,4], Shun-Qing Shen[5], Yong P. Chen[6,7,8] and Faxian Xiu[1,2,9*]

[1]State Key Laboratory of Surface Physics and Department of Physics, Fudan University, Shanghai 200433, China

[2]Collaborative Innovation Center of Advanced Microstructures, Nanjing 210093, China

[3]Institute for Quantum Science and Engineering and Department of Physics, South University of Science and Technology of China, Shenzhen 518055, China

[4]Shenzhen Key Laboratory of Quantum Science and Engineering, Shenzhen 518055, China

[5]Department of Physics, The University of Hong Kong, Pokfulam Road, Hong Kong, China

[6]Department of Physics and Astronomy, Purdue University, West Lafayette, Indiana 47907, USA

[7]Birck Nanotechnology Center, Purdue University, West Lafayette, Indiana 47907, USA

[8]School of Electrical and Computer Engineering, Purdue University, West Lafayette, Indiana 47907, USA

[9]Institute for Nanoelectronic Devices and Quantum Computing, Fudan University, Shanghai 200433, China

*Corresponding author: Faxian Xiu,  E-mail: Faxian@fudan.edu.cn

**Abstract**

The introduction of topological invariants, ranging from insulators to metals, has provided new insights into the traditional classification of electronic states in condensed matter physics. A sudden change in the topological invariant at the boundary of a topological nontrivial system leads to the formation of exotic surface states that are dramatically different from its bulk. In recent years, significant advancements in the exploration of the physical properties of these topological systems and regarding device research related to spintronics and quantum computation have been made. Here, we review the progress of the characterization and manipulation of topological phases from the electron transport perspective and also the intriguing chiral/Majorana states that stem from them. We then discuss the future directions of research into these topological states and their potential applications.

# 1 Introduction

As the foundation of condensed matter physics, band theory can accurately describe the basic electronic properties of most crystalline materials with weak correlation. Considerable attention has been paid to the energy gap and dispersion of electronic states. However, the Berry phase, an important part of the wave function, has long been neglected. The discovery of the quantum Hall effect (QHE) in the 1980s demonstrates that the phase of the wave function also contributes to electronic states [1, 2]. To characterize the difference between quantum Hall (QH) states and ordinary insulator states, a topological invariant called the Chern number, which is the result yielded via the integration of the Berry flux in the Brillouin zone, was introduced [2-4]. The Chern number is always a discrete integer and gives rise to the quantization of the Hall conductivity [2]. Similar to the concept of topology in mathematics, physical systems with different topological invariants cannot be continuously transformed into each other by finite perturbations [3, 4]. At the boundary between topological nontrivial and trivial systems, the energy gap must vanish; otherwise, it is impossible for the topological invariant to change [3, 4]. Since the discovery of the QH states, many exotic symmetry-protected topological phases in insulators have been uncovered via the introduction of new topological invariants, such as the $Z_2$ invariant [5] and mirror Chern number [6], based on different symmetries.

One representative example of topological systems is the topological insulator (TI), which exhibits metallic surface states regardless of whether a finite band gap exists in the bulk [3, 4]. This scenario is beyond the scope of the classical band structure model, which distinguishes conductors and insulators in terms of gap size. These gapless surface states originate from the bulk-boundary correspondence at the interface, where certain topological invariant changes [5]. Presently, the most widely studied TIs are $Z_2$ TIs and topological crystalline insulators (TCIs), which are protected by time-reversal symmetry (TRS) and crystalline symmetry, respectively [5, 6]. The panels on the left side of Figure 1 show schematic drawings of the bulk and surface states in a typical three-dimensional (3D) strong TI. As in graphene, the energy-momentum dispersion relationship of the surface states is linear at low energies, forming a Dirac cone. Characterized by an odd $Z_2$ number, the surface states in a TI are Fermi circles enclosing an odd number of Dirac points [7, 8]. In the absence of spin degeneracy, the Dirac cone surface states present a helical spin texture [8, 9]. The surface states of a TI can be described by a two-dimensional massless Dirac model [10]

$$H = \gamma(\sigma_x k_y - \sigma_y k_x),$$

where $\gamma$ is a parameter related to the effective velocity, $\sigma = (\sigma_x, \sigma_y)$ denotes the Pauli matrices, and $k = (k_x, k_y)$ is the wave vector. The model describes a conduction band and valence band touching at the Dirac point. Without loss of generality, we focus on the positive energy part. The spinor wave function is written as

$$\psi_k(r) = \frac{1}{\sqrt{2}}\begin{pmatrix} 1 \\ -ie^{i\varphi} \end{pmatrix} e^{ik\cdot r}$$

with $\tan\varphi = k_y/k_x$. The spinor wave function describes the helical spin structure of the surface states, which can give rise to a $\pi$ Berry phase when an electron moves adiabatically around the Fermi surface [11]. The Berry phase is a geometric phase collected in an adiabatic cyclic process [12], and can be determined as follows [13]

$$\phi_b = -i\int_0^{2\pi} d\varphi \left\langle \psi_k(r) \left| \frac{\partial}{\partial \varphi} \right| \psi_k(r) \right\rangle = \pi.$$

Dirac points on opposite surfaces can be viewed as partners in terms of spin degeneracy to satisfy the fermion doubling theorem [3]. On the other hand, the bulk states of TIs are gapped insulators with band inversion [7, 8]. The dispersion near the band gap can be well described by the following (3+1)-dimensional Dirac equation with a finite mass term [10, 14],

$$H = vk\cdot\alpha + (m - Bk^2)\beta,$$

where the wave vector $k = (k_x, k_y, k_z)$ and $v$, $m$, and $B$ are model parameters. If $mB > 0$, the model describes a TI. $\alpha = (\alpha_x, \alpha_y, \alpha_z)$ and $\beta$ are 4×4 Dirac matrices that obey the relations

$$\alpha_i^2 = \beta^2 = 1, \ \alpha_i\alpha_j = -\alpha_j\alpha_i, \ \alpha_i\beta = -\beta\alpha_i,$$

where $i$ and $j$ cover $x, y, z$. The Dirac matrices can be expressed using the Pauli matrices $\sigma_x, \sigma_y, \sigma_z$, and the identity matrix $\sigma_0$

$$\alpha_i = \begin{pmatrix} 0 & \sigma_i \\ \sigma_i & 0 \end{pmatrix}, \ \beta = \begin{pmatrix} \sigma_0 & 0 \\ 0 & -\sigma_0 \end{pmatrix}.$$

Therefore, the low-energy quasiparticles of the bulk states are actually massive Dirac fermions. To date, 3D TIs have been realized in many weakly interacting systems, including $Bi_{1-x}Sb_x$, $(Bi/Sb)_2(Se/Te)_3$, and $Pb_{1-x}Sn_x(Se/Te)$, [3, 6, 8, 15-17] and in correlated systems such as the Kondo insulator $SmB_6$ [18-22]. Spin-orbit coupled massless Dirac fermions in the TI surfaces give rise to numerous exotic phenomena, such as the quantum spin/anomalous Hall effect [23-26], Majorana zero mode [27], spin-orbit torque [28] and topological magneto-electric effect [29, 30].

Despite the thorough study of topological insulating systems, a comprehensive description of topological phases in metals has not yet been achieved because of the complexity of closed Fermi surfaces in metals. Nevertheless, the topological phases in certain semimetals, whose Fermi surface contains only isolated nodes or lines at zero energy, have been intensively investigated in recent years [31-34]. The nodes or lines originate from the band crossings.[31, 32] In a 3D system, a band crossing formed by two non-degenerate bands can be shifted but not removed by small perturbations because of the protection offered by monopole charges in momentum space.[31, 32] Conversely, band crossings in graphene are easily disrupted by perturbations because additional crystalline symmetries are required to form band crossings in two-dimensional (2D) systems.[3] The dispersion near these crossings of non-degenerate bands takes the form of a Weyl fermion, a long-pursued massless relativistic particle in high-energy physics, with the half freedom of a Dirac fermion.[31] The Weyl fermion can be described as the massless limit of the Dirac equation, *i.e.*, $H = vp \cdot \alpha$, or, equivalently, in the Weyl representation, *i.e.*,

$$H = \begin{pmatrix} H_+ & 0 \\ 0 & H_- \end{pmatrix} = \begin{pmatrix} +vk \cdot \sigma & 0 \\ 0 & -vk \cdot \sigma \end{pmatrix}.$$

To enable the crossings of non-degenerate bands, a system cannot simultaneously possess inversion (*P*) and time-reversal (*T*) symmetry. Otherwise, all the bands will be at least doubly degenerate (known as Kramer degeneracy).[3] Hence, this kind of topological semimetal (TS), *i.e.*, Weyl semimetal, can be found only in either non-centrosymmetric or magnetic materials.[31, 35] The corresponding band crossing points (Weyl nodes) exhibit two different types of chirality (±1) and act as monopoles (sinks or sources) of the Berry curvature in the momentum space.[35] Here, we take one of the Weyl fermions as an example. For $H_+ = +vk \cdot \sigma$, the monopole charge $N_+$ can be calculated as an integral on the sphere $\Sigma$ that encloses the Weyl point, *i.e.*,

$$N_+ = \frac{1}{2\pi} \int_\Sigma dS \cdot \Omega = -1,$$

where the Berry curvature $\Omega \equiv \nabla \times A = \mp 1/2k^2$ and the Berry connection $A \equiv -i\langle\psi_+|\nabla_k|\psi_+\rangle$ can be determined based on one of the eigen states $|\psi_+\rangle = [\sin(\theta/2), -\cos(\theta/2)e^{i\varphi}]^T$ of $H_+$. Similarly, for $H_- = -vk \cdot \sigma$, the monopole charge $N_- = 1$. Therefore, the Weyl points always exist in pairs with a net chirality of zero, as shown in the right panel of Figure 1. If the Weyl nodes with opposite chirality overlap in the momentum space, the pair may be annihilated, opening an energy gap formed via band hybridization. If additional crystalline point-group symmetry is preserved, a Dirac semimetal state will be generated, with the Dirac nodes protected by the rotational symmetry, mirror reflection symmetry, or *PT* symmetry.[36-39] This state can be considered as two opposite copies of Weyl semimetals with superimposed Weyl nodes.[36] An important consequence of the Berry curvature monopoles is that the Chern number of the 2D cross-sectional Brillouin zone changes by ±1 across each Weyl node. Similar to the QHE, the boundary states emerge at cross-sections enclosing nonzero net chirality (*i.e.*, the internode interval of Weyl node pairs) and terminate at the Weyl nodes. Hence, the surface states of TSs present an open loop in momentum space, called Fermi arcs (Figure 1), which is also spin-polarized.[31, 40] The Fermi arcs on opposite surfaces of a TS are joined together via their connection to the same pair of Weyl nodes. This completes a round trip, known as a Weyl orbit.[41] TS states have been found in several material systems, such as Dirac semimetals Na$_3$Bi [36, 42]/Cd$_3$As$_2$ [37, 43, 44] and the inversion-asymmetric Weyl semimetal TaAs family [35, 45-47]. These Weyl nodes and the open Fermi arc states lay the foundation for substantial new topological phenomena beyond the previously studied topological insulating systems. Moreover, the quasiparticle analogy of Weyl fermions allows high-energy physics models to be tested and the nature of the elementary particles in different types of condensed matter to be explored.

The discovery of these exotic topological phases of matter has initiated a rush towards the comprehensive investigation of topological physics and potential applications. In this review, we discuss the electron transport properties of these topological states and compare the properties of 3D TIs and TSs in weakly correlated electron systems with respect to a general framework. Note that although several new topological node-line and triplet semimetal materials have been proposed [48, 49], we mainly discuss Dirac and Weyl semimetals as examples of TSs. This review is organized as follows. First, we present a general overview of how to distinguish the bulk and surface states phenomenologically via electron transport. Specifically, a new cyclotron orbit, *i.e.*, the Weyl orbit, formed by the coupling of Fermi arc pairs in the surface states of TSs is introduced. Second, we describe the unique signatures of TIs and TSs and analyze their relation to the internal topological properties. We then present several approaches to manipulating the topological states via the breaking of symmetries and construction of hybrid systems. Finally, we discuss directions for future research into exotic physical phenomena and the potential device research based on their topological properties.

## 2   Transport methods for detecting bulk and surface states

Since the penetration depth of surface states is usually very small, *i.e.*, within tens of nanometers, transport signals are dominated by the bulk states in macroscopic samples for most TIs with a relatively high Fermi level or TSs. The bulk states can therefore be directly detected by utilizing conventional transport approaches. However, this task becomes exceedingly difficult when considering the surface states because of the large bulk contribution to transport. Since the beginning of the research on TIs, separation of the surface states from the bulk states has been one of the biggest challenges. The difficulty arises because the bulk gap in TIs is formed via band hybridization in inverted-gap systems and is therefore very small. Most TI materials have a gap of no more than 0.3 eV. Thus, they are inevitably affected by thermally activated carriers [15]. Moreover, to ensure strong spin-orbit coupling for band inversion, they usually contain elements such as Se and Te. These elements often exhibit naturally occurring crystalline defects and may be easily oxidized[15]. Extensive efforts, such as performing chemical doping [50-55], increasing the surface-to-volume ratio [56-59], and exploiting the electrostatic gating technique [60, 61], have been devoted to solving these problems to help enhance the surface conduction component via the enlargement of the bulk band gap or the tuning of the Fermi level inside the bulk gap. In TS systems, because of the semimetal nature of the bulk bands, bulk conduction is intrinsically problematic when probing the surface states. However, these methods can still be used to minimize the bulk density of states by tuning the Fermi level to be around the Weyl points. To date, the use of nanostructures and gate voltages has been adopted to study the Fermi arc surface states in TSs [62-64]. These efforts have yielded limited success because of the lack of low-carrier-density nanostructures and insufficient gate tunability.

In transport experiments, a direct way of distinguishing between the bulk and surface states is to probe the dimensionality of the Fermi surface (Figure 2a) based on the quantum oscillations. Under an external magnetic field, electronic states are quantized into a series of Landau levels that are periodic in energy. Changing the magnetic field (altering the Landau level degeneracy) or tuning the carrier density (shifting the Fermi level) causes the density of states near the Fermi level to oscillate as a function of the field or carrier density. This oscillation of the density of states is reflected in the conductivity and magnetic susceptibility, corresponding to Shubnikov–de Haas (SdH) oscillations and de Haas–van Alphen (dHvA) oscillations, respectively. When sweeping the magnetic field $B$, the oscillations are periodic in $1/B$. Their frequency is given by $F = (\phi_0/2\pi^2)S_F$, where $\phi_0 = h/2e$ and $S_F$ is the extremal cross section of the Fermi surface. Because of the linear band dispersion, topological systems usually exhibit high carrier mobility and make it relatively easy to detect quantum oscillations. By tracking the angle dependence of the oscillation frequency, the bulk and surface states can be distinguished by considering the Fermi surface dimensionality. Examples of this method can be found in previous studies of TIs such as $Bi_2Te_3$ [65]. Since the bulk Fermi surface may also be quasi-two-dimensional in certain layered systems, a more rigorous approach is to compare the Hall carrier density with the obtained Fermi surface size at the same time to help determine the bulk state. In low-Fermi-level samples, one can also use a high magnetic field to drive all the bulk carriers to the lowest Landau level, after which the remaining quantum oscillations are contributed solely by the surface states [66]. In TSs, a similar method is also used in investigating the Weyl orbit of $Cd_3As_2$ [64], a novel cyclotron orbit combining the Fermi arcs on two opposite surfaces and the bulk chiral Landau level [41]. Distinct from the surface states of TIs, the Weyl orbit acquires an additional quantum phase from the bulk propagation process [41, 67], therefore exhibiting a unique behavior related to the sample thickness. The analysis of SdH oscillations can provide important information regarding the electronic states, including the Fermi wave vector, effective mass, and quantum life time. Moreover, the combination of chemical doping or electrical gating with the SdH oscillations has been widely used during the transport measurements in topological systems [57, 68-70]. Dual-gate structures allow the Fermi levels of the top and bottom surface states to be independently controlled [71]. The energy-momentum dispersion relationship of the Dirac spectrum can be determined by analyzing the energy dependence of SdH oscillations.

Because of the 2D characteristic of surface states, several methods for detecting the surface transport in addition to probing the Fermi surface also exist. Among them, one of the most definite pieces of evidence for 2D transport is the QHE, in which the 2D electron gas, subjected to a perpendicular magnetic field and low temperature, forms a ballistic edge transport, leading to a vanishing longitudinal resistance and a quantized Hall resistance [1, 72] (Figure 2b). The Hall conductance $\sigma_{xy}$ is quantized to $ve^2/h$, where the filling factor $v$ represents the number of filled Landau levels, regardless of the sample details [1, 72]. Measurement of the Hall conductance offers a straightforward and precise way to determine the number of propagating edge modes. By tracking the thermal activation behavior of longitudinal resistance, the Landau level gap can also be obtained. Another common approach is to study the quantum interference behavior of conducting electrons in transport, such as Aharonov-Bohm (AB) oscillations and weak localization (WL)/ weak antilocalization (WAL). As illustrated in Figure 2c, the AB oscillations are resulted from the electrons traveling along paths that enclose magnetic flux in real space, inducing periodic oscillations of the conductance as a function of the flux [73]. The magnitude of the oscillations is a function of $B$, with a period of $h/e$. For example, an ideal TI nanowire (or nanoribbon), with a conducting surface and an insulating bulk, can be regarded as a hollow metallic cylinder to host this quantum interference.[56, 57] When the bulk states are sufficiently conducting, these oscillations are eliminated because of the destruction of a well-defined loop structure. Observation of these interference-induced oscillations provides strong evidence for surface transport. Normally, AB oscillations occur in the ballistic or pseudo-diffusive transport regime [73]. And it will evolve into Altshuler-Aronov-Spivak (AAS) oscillations with double the period ($h/2e$) during diffusive transport (Figure 2d) [73]. This is the result of the interference between clockwise and counterclockwise time-reversal loops, which doubles the accumulated magnetic flux [73]. Similar to the AAS oscillations from the conducting loops with multiple connections, the quantum interference of time-reversal loops in a flat geometry (Figure 2e) contributes an additional correction to the overall magnetoconductivity [74]. In conventional disordered metals, the quantum interference suppresses the conductivity as the temperature decreases, leading to WL, which is the precursor of Anderson localization [75]. In contrast, WAL often occurs in systems with strong spin-orbit coupling, manifesting itself as a positive correction to the conductivity [76]. We note that although WAL and WL are not exclusive to 2D systems, they can be useful in detecting the spin texture of surface states, which will be discussed later.

## 3 Tracking topological characteristics in transport experiments

### 3.1 Band crossing point

Having summarized several useful ways to probe the bulk and surface states, we now discuss how to utilize these methods to investigate the unique characteristics of these topological systems. The most direct consequence of a Dirac-type band is that of linear energy-momentum dispersion relationship, which is different from that of conventional quadratic bands, such as 2D electron gas. The linear band dispersion gives rise to many exotic properties such as high mobility and a large magnetoresistance (MR) [77-80]. Comparison of the related transport parameters, such as effective mass and mobility, can be a convenient way to help determine the dispersion near the Fermi level. However, in real materials, these Dirac/Weyl equations usually hold only for a low energy limit near Dirac/Weyl nodes, while the dispersion relation will not be strictly linear far away from the nodes because of some high-order corrections [3, 4, 40]. On the other hand, some other trivial systems, such as narrow-gap semiconductors, may also exhibit quasi-linear dispersion with high mobility [81]. Therefore, the band dispersion near the Fermi level functions more as a conclusive one for topological systems.

One of the most distinct features in a Dirac system is the existence of a band crossing point. In a magnetic field, the Landau quantization of 2D Dirac fermions can be expressed as

$$|\epsilon_N| = \sqrt{2|NB|\hbar e v_F^2 + (g\mu B/2)^2},$$

where $v_F$ is the Fermi velocity, $B$ is the magnetic field, and $g$ is the g-factor [4, 73]. Ignoring the Zeeman term yields $\epsilon_N^2 = 2|NB|\hbar e v_F^2$. In SdH oscillation measurements, where the magnetic field varies while the Fermi level is fixed, the critical field $B_N$, in which each Landau level passes the Fermi level, follows $N \propto 1/|B_N|$, as shown in Figure 3a. In comparison, in conventional quadratic bands, $\epsilon_N$ is equal to $(N + \frac{1}{2})\hbar\omega_c$, where the cyclotron energy $\omega_c \propto B$, thus giving $(N + \frac{1}{2}) \propto 1/B_N$. Therefore, by plotting the Landau fan diagram, the Dirac and quadratic dispersions can be distinguished using the phase shift $\gamma$. When generalizing to the 3D case, one needs to consider the dispersion along the third direction, which gives an extra dimensional coefficient to the phase shift (-0.125 for electron carriers and 0.125 for hole carriers) [82, 83]. The very existence of the phase shift in SdH oscillations is due to the Berry phase acquired from the cyclotron motion around a Dirac point in momentum space [84, 85]. Normally, this phase shift is perturbatively stable in the presence of a small gap term as confirmed via numerical simulations based on the gapped Dirac model.[86] However, if the band dispersion is not perfectly linear, the Berry phase becomes more sensitive to the gap term [86, 87]. Calculations show that in the semi-classical regime (large *N*), $\gamma$ varies between 0~1 in the presence of the gap term and high-order corrections [86]. In TSs, a Lifshitz transition, during which the Fermi surfaces enclosing opposite Weyl nodes merge, will occur, resulting in Landau level crossing at the band curvature [88]. Therefore, the Berry phase will be Fermi-level-dependent, changing from the nontrivial phase near the Weyl nodes to the trivial phase above the Lifshitz energy, with certain non-monotonic behavior near the Lifshitz transition [88]. This method of extracting the Berry phase is very helpful and straightforward for materials with a simple band structure. For multiband oscillations with strong interference between them, restoration of the Landau level positions for each band is a difficult process.

An important property of the band crossing point given by the Landau quantization equation mentioned above is that the zeroth Landau levels are fixed at zero energy, independent of the magnetic field (ignoring the Zeeman effect, Figure 3a) [15, 73]. These Landau levels are shared by the conduction and valence bands (*i.e.*, the higher and lower parts of the Dirac cone). The particle-hole symmetry guarantees that $\sigma_{xy}$ is an odd function of energy across the Dirac point [89]. The Hall conductance is $-e^2/2h$ when the Fermi level is below the zeroth Landau level; and jumps to $e^2/2h$ after the Fermi level is above the zeroth Landau level. Hence, a half-integer QH conductance $((n+\frac{1}{2})e^2/h)$ arises in the Dirac system. In the presence of small perturbations that can gap the band crossing points, the zeroth Landau levels will be shifted towards either the positive or the negative energy part, with the sign determined by the coupling term with perturbation, e.g., the two kinds of valleys in graphene with high magnetic field and TCI $Pb_{1-x}Sn_xSe$ bulk states, or the Zeeman energy shift in the partner surfaces of TIs [3, 10, 14, 89-91]. But the overall numbers of these two cases should be equal to meet the TRS. Therefore, the total density of states in the zeroth Landau levels should be half of that of the other Landau levels (Figure 3a), because of the energy splitting in spin or pseudospin (or valley) degree of freedom. This half-integer QHE has been indirectly observed in graphene (where with a four-fold degeneracy, a $(4n+2)e^2/h$ Hall conductance has been observed), which is a prototype Dirac system [92, 93]. In TI surface states, because of the partner surfaces, such as the top and bottom surfaces in thin film geometry, the overall Hall conductance will be quantized to a series of integer quantum values. For example, as a result of tuning the Fermi level inside the bulk gap, the QHE from the TI surface states has been observed, with a spacing of $e^2/h$, in Hall plateaus (Figure 3b) [68, 94]. This occurred because only the bottom surface Hall conductance was switched by the bottom gate voltage, while the top surface remained unaffected due to the electrostatic screening effect. Further experiments involving dual-gate devices also confirmed this scenario [71]. For 3D systems, one way to identify the degeneracy is to drive all the carriers to the lowest Landau level. By tracking the critical field $B^*$, in which the Landau level crosses over from *N*=1 to *N*=0, one can determine the carrier density via the calculation of $n = \sqrt{2}g_s g_L/(\pi l_B)$ by integrating the density of states of the Dirac cone. Here, $g_L$ and $l_B$ are the Landau level degeneracy per spin and the magnetic length, respectively. Comparison with the Hall carrier density can yield the spin/pseudospin degeneracy $g_s$ of the zeroth Landau level [91], which is strikingly different from conventional gapped bands by a factor of 2. Note that this method can be used only when the Hall effect is solely contributed by the carriers from the Dirac cone. Essentially, this method is somewhat equivalent to the Landau fan diagram approach, as both methods detect the Landau level position shift induced by the band crossing point.

### 3.2 Spin-momentum-locking boundary states

One of the most intriguing properties in topological systems is the spin-momentum-locking boundary states. They give rise to many geometric effects from the helical spin texture in surface/edge transport [73, 95]. As introduced in the transport method section, the quantum interference experiments are sensitive to the spin texture. Such helical spin texture contributes a Berry phase of $\pi$ to the electronic wave function, which strongly suppresses the backscattering from disorder and impurities. Therefore, the AB oscillations in TIs are found to be very robust due to the suppression of backscattering [56, 57, 73, 96]. In TI nanowires, the confinement of the surface along the transverse directions results in discretely quantized momentum and generates a series of one-dimensional sub-bands. The sub-band energy, *i.e.*, $\epsilon(k_\parallel) = \pm\hbar v_F \sqrt{k_\parallel^2 + k_\perp^2}$ with $k_\perp = 2\pi(l + 0.5 - e\Phi/h)/C$, can be tuned using the magnetic flux ($\Phi$) and the Fermi wave vector.[96] Here, 0.5 originates from the $\pi$ Berry phase, and $C$ is the circumference. By tuning the quantum phase, one can switch between a gapped mode and a gapless one in the AB oscillations (Figure 3c). Such a tunable helical mode has been observed in the surface states of $Bi_2Te_3$ nanoribbons [97]. The AB oscillation phase can be modulated by the Fermi wave vector using the gate voltage as well as the magnetic flux from the external magnetic field (Figure 3d), acting as a unique signature for the topological surface states.

This kind of quantum interference can also be extended to the diffusive transport regime where the AB oscillations are replaced with AAS oscillations for the nanowire/nanoribbon geometry [98]. The signature of the AAS oscillations has been found via fast Fourier transform (FFT) analysis of the conductivity oscillations [56, 57]. However, it is difficult to distinguish the AAS oscillations from the harmonic frequency of AB oscillations if the AAS oscillation amplitude is smaller than the AB oscillation amplitude in the FFT spectrum. A more common example of the diffusive quantum interference effect is the WAL behavior in TI thin films. One of the merits of TIs is the topologically protected surface states which cannot be Anderson-localized [3, 73, 99]. Because of the spin-momentum locking, the time-reversal loops in the TI surface states form destructive interference, leading to a delocalization behavior, in contrast to conventional disordered metal. A signature of this delocalization tendency is WAL [76]. This effect requires TRS. Therefore, a magnetic field can destroy it, resulting in a negative magnetoconductivity. The field dependence of the magnetoconductivity can be described by the Hikami-Larkin-Nagaoka formula [76] $\Delta\sigma = \alpha \frac{e^2}{\pi h}[ln\left(\frac{B_\phi}{B}\right) - \Psi(\frac{1}{2} + \frac{B_\phi}{B})]$, where $B_\phi = \hbar/(4De\tau_\phi)$ is a characteristic field related to the dephasing time $\tau_\phi$, $\alpha$ is a coefficient typically used to measure the spin-orbit coupling, and $\Psi(x)$ is the digamma function. The negative magnetoconductivity has been observed in almost every TI sample [56, 59-61, 100-104], and is widely considered to be the signature of the delocalization tendency of topologically protected surface states. For one conducting channel with spin-momentum locking, the value of $\alpha$ should be 0.5 in theory [76]. Because of the bulk contribution and the surface-bulk coupling, the obtained value of α in experiments varies from 0.3~1.2. Meanwhile, the conductivity provided by WAL is expected to increase with increasing sample size or decreasing temperature. However, in most experiments, a suppression of the conductivity with decreasing temperature has been observed below 10 K in $Bi_2Se_3$ and $Bi_2Te_3$ [101, 103, 104], showing a tendency towards Anderson localization [73, 75]. This dilemma faced in transport experiments has been found to be associated with the destruction of the super-metal picture of TIs due to electron-electron interactions [104], which was not previously taken into consideration [105]. The interplay between interaction and disorder suppresses the density of states on the Fermi surface, leading to a temperature dependence of the conductivity similar to that of WL [106, 107], known as the Altshuler-Aronov-Fukuyama (AAF) effect. A Feynman diagram calculation, based on the massless Dirac model, shows that the surface states of a TI are not immune to the AAF effect [108]. The AAF effect leads to a conductivity contribution that decreases logarithmically with decreasing temperature in the presence or absence of a magnetic field, which overwhelms the WAL regarding the temperature dependence of conductivity. By tuning the electron interaction strength in the TI nanostructures with antidots, the trend of the temperature dependence of conductivity has been found to be consistent with the theory of interaction-induced localization [109].

In contrast to these exotic phenomena discovered in TI surface states, the progress of TS surface states in transport experiments is limited. As of now, AB oscillations have been observed only in Dirac semimetal $Cd_3As_2$ nanowires [63]. An anomalous phase shift in the AB oscillations has been found under a high magnetic field (over 4 T). This phase has been interpreted as the field-induced separation of Weyl nodes, which influences the overall AB phase. The study of TS surface states is still in its early stage, with many mysteries waiting to be clarified, especially the exotic Weyl orbit with interplay between real space and momentum space. Comparison of the various surface-related phenomena observed in TSs to those of TIs and studies on the role of different surface states in these effects are also worthwhile.

### 3.3 *Novel orbit formed by surface Fermi arcs and bulk chiral model*

In conventional systems with closed Fermi surfaces far away from the Brillouin zone boundary, electrons will be driven to travel periodically along a closed path in momentum space (cyclotron orbit) by the Lorentz force in an external magnetic field. Normally, the change in momentum is continuous, with no kink or gap throughout the orbit. However, this concept no longer holds when generalized to the Fermi arc surface states in TSs. As the most unique characteristic of TS surface states, a Fermi arc is an open loop that connects a pair of Weyl nodes without reaching the boundary of the Brillouin zone. For each $k_z$ between the Weyl nodes, the system is equivalent to a 2D topological Chern insulator. There are edge states at a surface parallel to the direction along which the Weyl nodes separate, *e.g.*, the edge states exist at four surfaces $y$=0, $L$ and $x$=0, $L$, if the Weyl nodes are separated along the $z$ direction. The energy spectrum of all the edge states forms a tilted disk in momentum space [110]. The intersection between the disk and the Fermi energy forms the Fermi arc. Under a magnetic field, electrons travel to the end of the Fermi arc, *i.e.* the Weyl nodes, but cannot complete a cyclotron motion within the surface states. However, Potter *et al.* found that the gapless chiral mode in bulk bands allows electron cyclotron motion to extend beyond the Fermi arc [41]. This chiral mode disperses in the direction of the field, propagating towards the opposite surface, and transfers electrons between Weyl nodes at different surfaces. Upon reaching the opposite surface, the electrons cycle through the corresponding Fermi arc and return to the initial surface via the bulk states, forming a closed path (a Weyl orbit) as shown in Figure 3e. The Lorentz force is ignored in the bulk electron transfer process because the momentum is parallel to the magnetic field. Therefore, the electrons can successfully reach the opposite surface without cycling in the bulk. This scenario also implies that electrons cannot be scattered and made to drift away from their initial path in the bulk. Hence, the thickness of the system should be smaller than the mean free path.

By considering the time required for electrons to travel through the surface and bulk, the effective Landau level energy can be expressed as $\epsilon_n = \frac{\pi v(N+\gamma)}{L+k_0 l_B^2}$, where $N$ is the Landau level index, $\gamma$ is a constant of order unity that encodes low-$n$ quantum effects, $k_0$ is the length of the Fermi arc, $L$ is the thickness, and $l_B$ is the magnetic length [41]. By further considering the phase-space quantization [67], the corresponding magnetic field $B_N$, at which the $N$th Landau level crosses the Fermi level $E_F$, can be determined using $\frac{1}{B_N} = \frac{2\pi e}{S_k}[n + \gamma - \frac{L}{2\pi}(k_{w\parallel} + 2k_{F\parallel})]$, where $k_{w\parallel}$ and $k_{F\parallel}$ are the component of the k-space separation of the Weyl node pair and the Fermi wave vector along the magnetic field direction, respectively. This equation indicates that the Weyl orbit has a periodic oscillation that is a function of $1/B$. The effective Fermi surface is the area enclosed by two Fermi arcs when projected onto the same surface. The Fermi surface size can be estimated using $F_S = \frac{E_F k_0}{e\pi v}$. For a coherent oscillation to form, a consistent field $B_N$ should be applied across the sample for each Landau level. To achieve this, the sample thickness must be uniform. Hence, a triangular, in contrast to a rectangular one, should not exhibit these oscillations (Figure 3f). This difference can be used to distinguish a Weyl orbit from conventional surface states. Under a large magnetic field, the oscillation may exhibit non-adiabatic effects resulting from shifts in the oscillation position caused by electrons jumping between Fermi arcs and bulk chiral modes. This behavior is also a unique signature of the Weyl orbit, and was successfully observed by Moll *et al.* in $Cd_3As_2$[64].

### 3.4 *Chiral anomaly of Weyl fermion*

The bulk of a TS is described by the massless Dirac equation and can be viewed as the quasiparticle analogy of Weyl fermions [40]. In contrast to a 2D Dirac system, the Landau level spectrum of 3D TSs includes gapped non-chiral bands ($N>0$), and gapless chiral modes ($N=0$) as shown in Figure 4a, because of the dispersion along the magnetic field direction. The zeroth chiral Landau level of each Weyl cone disperses in only one direction with the sign determined by the chirality $\chi$, as given by the equation $\epsilon_0 = -\chi \hbar v_F k_\parallel$. Here, if an electric field **E** is applied parallel (or anti-parallel) to the magnetic field **B**, electrons will move in the opposite direction with respect to **E**. The anomaly occurs at the chiral Landau levels at which electrons are transferred from the right-moving band to the left-moving one (or in an opposite manner for the anti-parallel case) due to their one-way dispersion nature [111, 112]. Note that during this process, the electron number is conserved since the Weyl nodes come in pairs. This chirality-dependent charge pumping rate can be derived as $\frac{\partial \rho_\chi}{\partial t} = \chi \frac{e^2}{4\pi^2 \hbar^2} \mathbf{E} \cdot \mathbf{B}$, with $\rho_\chi$ being the pumped charge number. The system achieves an equilibrium state with non-equal charge occupation at two chiral modes after relaxation via inter-valley scattering, resulting in a chiral chemical potential difference between two Weyl cones. Subsequently, an axial current is induced in the direction opposite to that of the electric field because of the nonzero net momentum. Although initially proposed to be in the quantum limit [112], the chiral anomaly was later extended to the semi-classical regime [113]. Experimentally, one direct consequence of the chiral anomaly is the negative longitudinal MR induced by the axial current. The conductivity correction produced by the chiral anomaly is predicted to be proportional to $B^2$ in the low magnetic field regime [113]. This negative longitudinal MR has been widely observed in many TS systems such as ZrTe$_5$/HfTe$_5$, Nb$_3$Bi (Figure 4b), Cd$_3$As$_2$ and TaAs [62, 114-123]. However, extreme caution must be taken in distinguishing the chiral anomaly from other trivial effects. For example, current jetting will induce a voltage drop between the two inner electrodes in a four-terminal measurement because of the inhomogeneous current distribution. This voltage drop manifests itself as a negative MR [118]. The current jetting effect often occurs in high-mobility systems. In most TSs, the negative MR originating from the chiral anomaly is mixed with, or even overwhelmed by, the conventional positive MR. Hence, it is difficult to quantify these MR responses. Because the chirality-polarized state exhibits diffusion similar to that of spin polarization, a nonlocal transport experiment, in which the two resistance correction terms can be separated, was proposed to measure the diffused component of the chiral anomaly [124] (Figure 4c). This proposal was later realized in Cd$_3$As$_2$ nanoplates (Figure 4d) [119]. The nonlocal resistance showed a magnetic field dependence opposite to that of the local resistance and decreased exponentially with the diffusion length. A relatively long valley relaxation length of up to 2 μm was obtained from the experiment. Although the conductivity correction of the chiral anomaly is independent of $\mathbf{E}$ at the zeroth order, the higher-order response of the time derivative of momentum becomes non-negligible in a semi-classical Boltzmann transport equation because of the chiral chemical potential difference (a function of $\mathbf{E} \cdot \mathbf{B}$) [125]. The response leads to a nonlinear conductivity when the magnetic field is parallel to the current, which was recently observed in Bi$_{0.96}$Sb$_{0.04}$ [125].

## 4  Breaking TRS in topological phases

Introducing magnetization to a TI system is known to gap the Dirac cone on the surface that is perpendicular to the magnetization vector, as the helical topological surface states require TRS[3]. In a 3D TI film with perpendicular magnetization, the gapped Dirac states on opposite surfaces have different topological numbers due to the opposite magnetization direction of each surface [24]. The most interesting consequence of magnetizing the TIs is the emergence of chiral edge states in thin films, which carry a quantized Hall conductance of $e^2/h$ per channel, known as the quantum anomalous Hall effect (QAHE) [24, 95, 126]. Despite the significant differences between the integer QHE and the QAHE in material details and the way of breaking TRS, the underlying physics both belong to the scope of Chern insulators, with equivalent topological origins [95]. A comparison of the QAHE and the quantum spin Hall effect (QSHE) is presented in Figure 5a. The QSHE occurs in 2D TIs that have a gapped bulk state and gapless edge state [23]. It can be regarded as the doubling of the QAHE in the sense that a 2D TI has an even number of edge states with pairs of counter-propagating modes due to the preserved TRS [95]. The edge states in 2D TIs are basically the 2D versions of the helical Dirac surface states of 3D TIs. These three Hall effects all support dissipationless transport. However, unlike the QHE, the QAHE does not require an external magnetic field and is more robust than the QSHE because of the absence of scattering between two counter-propagating edge modes.[95] To gain further insight into the QAHE, the minimal model for the QAHE in a magnetic TI thin film is expressed by [127]

$$H = \left(\frac{\Delta}{2} - Bk^2\right)\tau_z \otimes \sigma_z + \gamma\tau_0 \otimes (\sigma_x k_y - \sigma_y k_x) + V\tau_x \otimes \sigma_0 + \frac{m}{2}\tau_0 \otimes \sigma_z,$$

where $k_x$ and $k_y$ are the wave vectors, $\tau$ and $\sigma$ are Pauli and unit matrices, respectively, $\Delta$ and $B$ are parameters influenced by the finite-size effect [128], $\gamma$ is the effective velocity, $V$ measures the structure inversion asymmetry (SIA) between the top and bottom surfaces of the thin film, and $m$ describes the exchange field of the magnetic dopants. When $|m| > \sqrt{\Delta^2 + 4V^2}$, *i.e.*, the exchange field overcomes both the finite-size gap and SIA, the system enters the QAHE phase.

Regarding the way to construct the quantum anomalous Hall state in a TI, Yu *et al.* proposed doping the $Bi_2Se_3$ family with magnetic impurities to develop magnetic order [24]. According to their mean field calculations, the magnetic impurities are coupled via the strong Van Vleck mechanism, thereby producing a ferromagnetic state even in a bulk insulating case without any itinerant carriers. However, another mechanism resulting in ferromagnetism, *i.e.*, the carrier-mediated Ruderman-Kittel-Kasuya-Yosida (RKKY) interaction, may also occur, as frequently observed in dilute magnetic semiconductor systems [129]. In experiments, different scenarios in which the ferromagnetic order can or cannot be tuned using the carrier density has been uncovered [130-133]. At present, a generally accepted hypothesis is that the Van Vleck mechanism is most significant in the bulk states of TIs while the RKKY interaction is dominant in the surface states [133, 134]. Once the ferromagnetic order has been established in an ultrathin TI film, the QAHE can be observed by tuning the Fermi level inside the bulk gap. However, substantial efforts are required to overcome this obstacle, as the magnetic doping is a two-sided process. The doping level changes the magnetic order as well as the carrier density. With a high doping concentration, a topological phase transition is further induced due to the reduced spin-orbit coupling strength.

In transport experiments, the signature of the induced surface gap upon TRS breaking can be probed via a crossover from WAL to WL. In the presence of the gap opening $\Delta$, the Berry phase of Dirac fermions becomes $\phi_b = \pi(1 - \Delta/2E_F)$, where $E_F$ is the Fermi energy measured from the Dirac point.[135] Thus, if the Fermi energy is not in the gap but crosses one of the bulk bands and is near the band edge, where $E_F = \Delta/2$, then the Berry phase becomes trivial, i.e., $\phi_b$=0 or $2\pi$. A $\pi$ Berry phase leads to WAL, and the trivial Berry phase indicates that the gap can drive a crossover from WAL to WL. Lu, Shi, and Shen developed the analytical formula of the magnetoconductivity to describe the WAL-WL crossover [135]

$$\Delta\sigma(B) = \sum_{i=0,1} \alpha_i [\psi\left(\frac{\ell_B^2}{\ell_\phi^2} + \frac{\ell_B^2}{\ell_i^2} + \frac{1}{2}\right) - \frac{\ell_B^2}{\ell_\phi^2} + \frac{\ell_B^2}{\ell_i^2}],$$

where $\psi$ is the digamma function, $\ell_B^2 \equiv \hbar/4|eB|$, and $\alpha_i$ and $\ell_i$ are sophisticated functions of the Berry phase. Basically, when $\phi_b = \pi$, $\alpha_1$=-1 and $\alpha_0 = 0$, then WAL occurs. As $\phi_b$ approaches 0, $\alpha_1 \to 0$ and $\alpha_0 \to 1$, giving rise to the WAL-WL crossover. $\ell_i$ also changes with the Berry phase but remains divergent when $\phi_b = \pi$ and 0, ensuring that the system is in a quantum diffusion regime.

If the Fermi energy is further tuned inside the surface gap, one can observe the quantized Hall conductance contributed solely by the edge states. Therefore, the crossover may serve as a precursor of the QAHE and has been observed in many related devices, such as Fe-doped $Bi_2Te_3$ [102], Cr-doped $Bi_2Se_3$ [136], Mn-doped $Bi_2Se_3$ [137], EuS/$Bi_2Se_3$ bilayers [138], and $Sb_{1.9}Bi_{0.1}Te_3$ on the ferrimagnet $BaFe_{12}O_{19}$ [139], and even in ultrathin $Bi_{1.14}Sb_{0.86}Te_3$ [140] due to the finite-size effect [128]. After extensive efforts to optimize the TI materials to achieve truly insulating bulk states, the QAHE was achieved in Cr- and V-doped $(Bi_{1-x}Sb_x)_2Te_3$ systems [26, 134, 141, 142] (Figure 5b). Recently, however, a new method for introducing magnetic order into TIs has been discovered via the proximity to ferromagnetic or antiferromagnetic materials [143-145]. Particularly, the anomalous Hall effect in the TI-antiferromagnetic insulator structure has been found to persist up to 90 K because of the robustness of antiferromagnetic coupling [145]. These studies demonstrate the significant progress made towards realizing the QAHE and dissipationless transport at high temperatures.

TRS breaking also yields many interesting physics in TSs. First, it is one of the two common ways to split the Weyl nodes and generate a Weyl semimetal phase. In the early stage of TS research, candidates of Weyl semimetals were mostly proposed based on the breaking of TRS [31, 33]. In TSs, breaking TRS via the magnetic order or an external field can also shift the Weyl nodes [36, 146]. A transition from Dirac to Weyl semimetal states resulting from Zeeman splitting has been suggested to occur under a magnetic field [36, 146] and the related experimental signatures have been observed [78, 80, 147]. Recently, half-Heusler GdPtBi was found to exhibit the signature of a chiral anomaly under a magnetic field [148]. In this material, the Weyl nodes are generated via band crossing due to the Zeeman effect. Similar ideas regarding the formation of Weyl nodes in Te [149] and $Pb_{1-x}Sn_xTe$ [150] systems via pressure-induced gap closing have also been proposed. However, one peculiar property of magnetic TSs is that the net Berry curvature from TRS breaking induces an anomalous Hall effect. For a pair of Weyl nodes located at $\pm k_w$ along the $k_z$ axis, the $k_x - k_y$ plane, with $|k_z| < k_w$, can be regarded as a 2D Chern insulator with a Berry flux flow in the $k_z$ direction [151-153]. Summing all the planes with nonzero Chern numbers yields a net anomalous Hall conductivity (Figure 5c). Here, the anomalous Hall conductance cannot be described by the $k \cdot \sigma$ model used to describe one individual Weyl node. To calculate the anomalous Hall conductance, one must adopt a two-node model [154], *i.e.*,

$$H = A(k_x \sigma_x + k_y \sigma_y) + M(k_w^2 - k_x^2 - k_y^2 - k_z^2)\sigma_z.$$

This model describes two Weyl nodes at $(k_x, k_y, k_z) = (0,0,\pm k_w)$. One can regard $k_z$ as a parameter and calculate a $k_z$-dependent Chern number by integrating the Berry curvature over the $k_x - k_y$ momentum space [128], as follows

$$n_c(k_z) = -\frac{1}{2}[\text{sgn}(M k_w^2 - M k_z^2) + \text{sgn}(M)].$$

The above expression indicates that for each $k_z$ between $[-k_w, k_w]$, one has a nontrivial Chern number, which contributes an anomalous Hall conductance of $\frac{e^2}{h}$. An integral of $n_c$ over $k_z$ then yields the anomalous Hall conductance $\frac{e^2}{\pi h}k_w$.[151] In TSs with TRS, the total anomalous Hall effect is zero since the $k_w$ values of different Weyl node pairs cancel each other out. In contrast, magnetic TSs will yield an anomalous Hall conductivity from the net Berry curvature. A large anomalous Hall/Nernst effect contributed by the Berry curvature has been reported for several TS materials or candidates such as GdPtBi [155], Mn$_3$Sn [156, 157], Cd$_3$As$_2$ [158] and Pr$_2$Ir$_2$O$_7$ [159]. The most notable feature is that the anomalous conductivity has a topological origin and is no longer directly coupled with the magnetization strength. For example, in Mn$_3$Sn [157] and GdPtBi [155], a relatively small magnetization or Zeeman energy can induce a large Berry curvature, which is beyond the scope of the conventional anomalous Hall/Nernst effect in ferromagnets (Figure 5d).

## 5 Manipulating topological systems for device applications

The successful manipulation of topological surface states and the generation of chiral electrons make it possible to develop application-oriented devices. Two major directions are considered to be promising in regard to utilizing these topological states: (1) topological quantum computation based on the robustness of the topological property and (2) the exploitation of the spin-momentum locking properties by spintronic devices. Here, we give an overview of recent progress, focusing on current achievements made regarding TI systems.

Deeply rooted in quantum mechanics, quantum computation has become a basis for future data operation but remains infeasible because of the fragile coherence of qubits and high error rates.[160] One solution to the latter problem is to conduct topological quantum computation using non-Abelian statistics.[161] The fault tolerance of this approach arises from the nonlocal encoding of the states, making them immune to errors caused by local perturbations. In addition to the fractional QH and anyon states, topological superconductivity has emerged as a new way of implementing non-Abelian statistics[4]. The consequent p-wave superconducting pairing results in a non-Abelian Majorana zero mode at the end of a 1D system or in the vortex core in a 2D system [4, 27]. Several proposals for constructing the Majorana zero mode have been suggested, including the use of odd-parity Cooper pairs and superconducting topological materials[4]. All the Majorana proposals share a similar mechanism, *i.e.*, the construction of a *p+ip* superconductor by introducing s-wave pairing into a system with spin-orbit interaction and Zeeman splitting. This mechanism can be explained via a two-dimensional example. The minimal Hamiltonian [162] can be written as $H = H_0 + V$, where

$$H_0 = \sum_{k,\sigma} c_{k,\sigma}^\dagger [\epsilon(k)\sigma_0 + \alpha(k_x\sigma_y - k_y\sigma_x) + V_z\sigma_z]_{\sigma,\sigma'} c_{k,\sigma'},$$

$$\text{and } V = \sum_k (\Delta c_{k,\uparrow}^\dagger c_{-k,\downarrow}^\dagger + h.c.).$$

Here, $\epsilon(k) = k^2/2m - \mu$, $\mu$ is the chemical potential and $\alpha$ reflects the Rashba spin-orbit coupling. The Hamiltonian can be transformed into two modified Dirac equations that describe two spinless $p_x \pm ip_y$ wave pairing superconductors. Near $k = 0$, the $+$ branch is trivial because its Chern number is always 0. For the $-$ branch, the Hamiltonian is expressed as

$$H \simeq \frac{1}{2}\sum_k \psi_{k,-}^\dagger [(\sqrt{\mu^2 + \Delta^2} - |V_z|)\sigma_z - (\alpha\Delta/|V_z|)(k_x\sigma_y + k_y\sigma_x)]\psi_{k,-},$$

and its Chern number can be found as follows

$$n_c = \frac{1}{2}\left[\text{sgn}(\sqrt{\mu^2 + \Delta^2} - |V_z|) - 1\right],$$

which explicitly indicates that this $p_x + ip_y$ superconductor is topologically nontrivial if $\sqrt{\mu^2 + \Delta^2} < |V_z|$, *i.e.*, the Zeeman energy must overcome the chemical potential and the s-wave pairing to induce the topological phase transition.

In transport, a notable feature of the Majorana zero mode is the zero-bias conductance peak (ZBCP), as initially observed in InSb/InAs nanowires coupled with superconducting electrodes (indicated by the dashed oval in Figure 6a) [163-165]. A magnetic field is required to induce the topological state via the Zeeman effect. The ZBCP has also been detected via scanning tunneling microscopy, along with real-space images confirming that the Majorana zero mode localizes at the ends of a 1D atom chain [166]. ZBCPs have been observed in several topological materials, such as $Cu_xBi_2Se_3$ [167], the $Bi_2Te_3/NbSe_2$ heterostructure [168], and $Cd_3As_2$ and TaAs with a hard point contact [169-171], either coupled with superconductors or intrinsically with superconductivity. However, ZBCPs may also be induced by other factors such as Kondo effect or disorder [172]. One way to distinguish them is the detection of the spin selectivity via Andreev reflection [173, 174]. Another proof of the Majorana fermion is the 4π-periodic Josephson effect [175]. While Cooper pairs with a charge of 2*e* can tunnel in the conventional Josephson effect, Majorana fermions enable the tunneling of a single electron, leading to the doubling of the periodicity in the Josephson effect. This phenomenon has been observed in the InSb/Nb nanowire junction [176] and strained HgTe/Nb junction [177]. In Figure 6b, an anomalous doubled Shapiro step (missing the *n*=1 step) appears at low frequency, consistent with the scenario of the Majorana bound state [177]. Apart from these reports of 0D Majorana modes resulting from the splitting of a single fermionic bound state, it has been proposed that an integer QH edge state can also be split into two 1D chiral Majorana modes in a spinless 2D *p+ip* superconductor [178]. The advantage of this propagating chiral Majorana mode is that it can form quantized transport, which is favorable for experimental detection. This proposal was recently realized in a QAH-superconductor heterostructure [179]. Half-integer quantized conductance plateaus appear when the magnetization is reversed, as shown in Figure 6c, serving as relatively strong evidence for the 1D Majorana mode [179], although other possible origins [180, 181] need further clarification. In the meantime, the realization of intrinsic topological superconductors is also very important for practical device applications. There has been significant interest in extending the progress made for superconducting hybrid topological systems to intrinsic topological superconductor candidates, such as $Cu_xBi_2Se_3$ [167], FeSe [182], and $PbTaSe_2$ [183].

Apart from the application to quantum computing, TIs are also suitable for use in spintronic applications because of their spin-momentum locking properties, which produce a spin-helical current in transport. Specifically, a directional electrical current (**I**) generates an in-plane net spin polarization along **n**×**I**, where **n** is the normal direction of the surface [28, 184]. The polarized spin will flip upon the reversal of the current direction or travel to the opposite surface. The net spin polarization can be probed directly by measuring the voltage on a ferromagnetic contact. The voltage is proportional to the projection of the spin polarization onto the magnetization direction of the contact [184]. Reversing the current or magnetization direction causes an abrupt change in voltage. A detailed analysis of the polarization direction and current density dependence can be carried out to distinguish the spin-momentum locking from the conventional Rashba effect [184]. The net spin polarization can diffuse into an adjacent conducting material, resulting in a pure spin current without a net charge flow [28]. The ratio between the obtained 3D spin current density and the applied 2D current density can be used as a quantitative index to evaluate the charge-to-spin conversion efficiency [185]. The spin current density can be detected based on the spin torque ferromagnetic resonance, where an *ac* charge current is used to generate the spin torque in an adjacent ferromagnetic layer (Figure 6d).[28, 185] The spin torque oscillates with the excitation current, inducing the procession of magnetization in the ferromagnetic layer, which can be detected based on its magnetoresistance. This spin-torque effect has been observed in TI-ferromagnetic metal [28, 185], and TI-ferromagnetic TI hybrid structures [186]. It can also be used to achieve magnetization switching through applying a current [186]. By tuning the Sb concentration in $(Bi_{1-x}Sb_x)_2Te_3$ thin films, the charge-to-spin conversion coefficient has been found to be nearly constant in the bulk insulating case but experiences a sharp minimum near the Dirac point (Figure 6e) [185]. This phenomenon may be caused by the instability of the helical spin structure near the Dirac point due to the scattering between Dirac fermions with opposite polarizations above and below the Dirac point. The robust spin-momentum locking of topological systems also enables the current-induced spin-polarization [28, 187] to persist up to room temperature. More interestingly, recent spin potentiometric measurements of $Bi_2Te_2Se$-based devices revealed an exceptionally long spin lifetime of up to several days at low temperature (Figure 6f) [188]. This long spin lifetime may be related to the coupling between nuclear spins and topological surface states. The efficient spin injection and long spin lifetime make TIs a very promising platform for spintronic applications.

# 6 Outlook

Significant advances in the study of topological states of matter have been achieved over the past decade. This field is likely to stay booming with the rapid accumulation of exciting new physics towards a transformative understanding of topological matters. By reviewing the overall transport properties of topological systems, one might realize that the exploration of the emerging surface states and device applications corresponding to TS systems are still premature compared with that for TI systems. It would be particularly beneficial, for example, to investigate the QHE and quantum interference with the additional Berry phase to better characterize the physics associated with the Weyl orbit in TSs. Weyl semimetals with broken TRS have also been proposed [31, 33], but have not yet been strictly confirmed. In these magnetic Weyl semimetals, the anomalous Hall effect due to the Berry curvature has been theoretically predicted and can be transformed into the QAHE by reducing the dimensionality [33]. Floquet-Bloch states, in which Bloch electrons interact with light, lead to new periodicities of electronic energy bands. Theory has predicted that illuminating Weyl semimetals with circularly polarized light may effectively break the TRS, allowing tuning of the Weyl nodes and Fermi arcs [189, 190]. Recently, type-II Weyl semimetals with tilted Weyl nodes have been predicted and confirmed in $WTe_2$ and $MoTe_2$ [191-196]. Because they violate Lorentz invariance, type-II Weyl fermions have no analogy in elementary particles. The tilted Weyl nodes result in an anisotropic chiral anomaly [191] and an unusual magnetoresponse of the Landau quantization [197]. The new cyclotron orbit of Fermi arcs is also predicted to host many exotic properties, including the QH effect with a 3D feature [198] and resonant transmission [199]. Particularly, QHE has been observed in the low-dimensional nanostructures and thin films of TS $Cd_3As_2$.[200-202] Further investigations are required to clarify whether it comes from the Weyl orbit with the z-direction propagation process or the deformed surface Dirac cones. Furthermore, correlation effects in topological phases are important in the study of TIs and TSs. By driving the system to the quantum limit, electron-electron interactions can be significantly enhanced, as demonstrated by several recent experiments on the phase transitions of TSs under high magnetic field [203, 204]. The interaction of Dirac electrons can generate a mass term to the band dispersion and induce electronic instabilities such as the spin density wave because of the peculiar spin texture [205]. On the other hand, strongly correlated materials such as Kondo insulators [206] and iridates [207] have been found to exhibit signatures of the topological phase. The interplay between the topological property and electron-electron interactions can offer new insights into superconductivity and fractionalized states.

Regarding the device applications based on topological systems, the aforementioned quantum computation and spintronics are still major directions. Exploration of the Majorana-associated phenomena, such as the anomalous Josephson effects, based on pristine superconducting topological systems such as $Cu_xBi_2Se_3$ [167], and the recently observed superconducting TS [183], would be interesting. Following numerous quantum computation proposals [160, 161], it is necessary to demonstrate basic qubit operations in topological systems by manipulating and reading the occupation of Majorana zero modes. Most importantly, an unambiguous demonstration of non-Abelian statistics using Majorana zero modes would be a significant step. In the aspect of spintronics applications, the realization of bulk insulating TI at room temperature is important for the utilization of spin transfer torque. Modulating topological systems with magnetic order at high temperatures, such as TI-antiferromagnetic insulator structures,[145] can be used to achieve dissipationless transport. Recent *ab initio* calculations demonstrated a strong intrinsic spin Hall effect in TaAs-based materials originating from the large Berry curvature and spin-orbit coupling of bulk bands.[208] Finally, the proposed voltage-driven magnetization switching in magnetically doped Weyl semimetals, may extend the application of TS systems to topological-related spintronics.[209] Multiple electronic devices utilizing the chirality of Weyl fermions have also been designed as possible platforms for valleytronics.[210, 211]

While the fundamental research on new topological states and the various emergent phenomena arising from them continues to be conducted, much attention has been paid to the potential utilization of topological properties in practice. Although significant gaps in our knowledge of these systems and their applications still exist, the progress made in recent years has been encouraging. The investigation of topological systems may far exceed our initial expectations, being just as amazing as it was at its initiation.

# Acknowledgements

F.X. was supported by the National Key Research and Development Program of China (Grant No. 2017YFA0303302), National Natural Science Foundation of China (Grant No. 11474058, 61674040). H.Z.L was supported by the National Key R & D Program (2016YFA0301700), the National Natural Science Foundation of China (11574127), Guangdong Innovative and Entrepreneurial Research Team Program (2016ZT06D348), and Science, Technology and Innovation Commission of Shenzhen Municipality (ZDSYS20170303165926217). S.Q.S. was supported by Research Grants Council,

University Research Committee, Hong Kong under Grant No. 17301116 and C6026-16W. We thank Awadhesh Narayan, Xiang Yuan, and Yanwen Liu for helpful discussions.

## Author contributions

C.Z., H.L., and F.X. drafted the manuscript and all the authors contributed in finalizing the manuscript.

## Competing financial interests

The authors declare no competing financial interests.

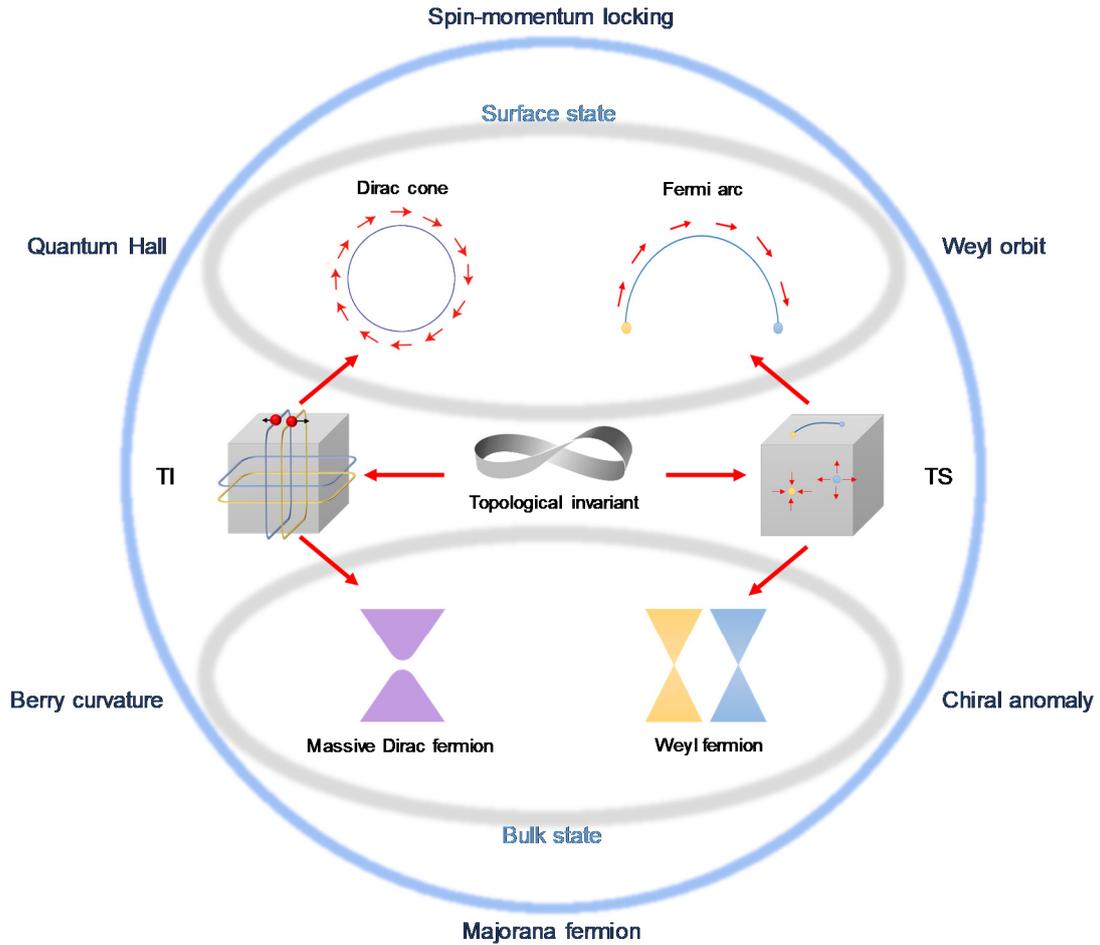

**Figure 1 (Color online). Comparison of different topological systems and related phenomena.** A schematic illustration of the surface and bulk states of a TI and a TS. The bulk states of the TI and TS are massive Dirac fermions and Weyl fermions, respectively. The surface states of the TI and TS form a closed Fermi pocket with a helical spin texture and open Fermi arcs with spin-momentum locking, respectively. Several representative examples of the topological properties are listed, such as the Weyl orbit (Figure 3e-f), chiral anomaly (Figure 4), QH-related phenomena (Figure 3b, Figure 5a-b), Berry curvature (Figure 5), spin-momentum locking (Figure 3c-d, Figure 6d-f), and Majorana fermion (Figure 6a-c).

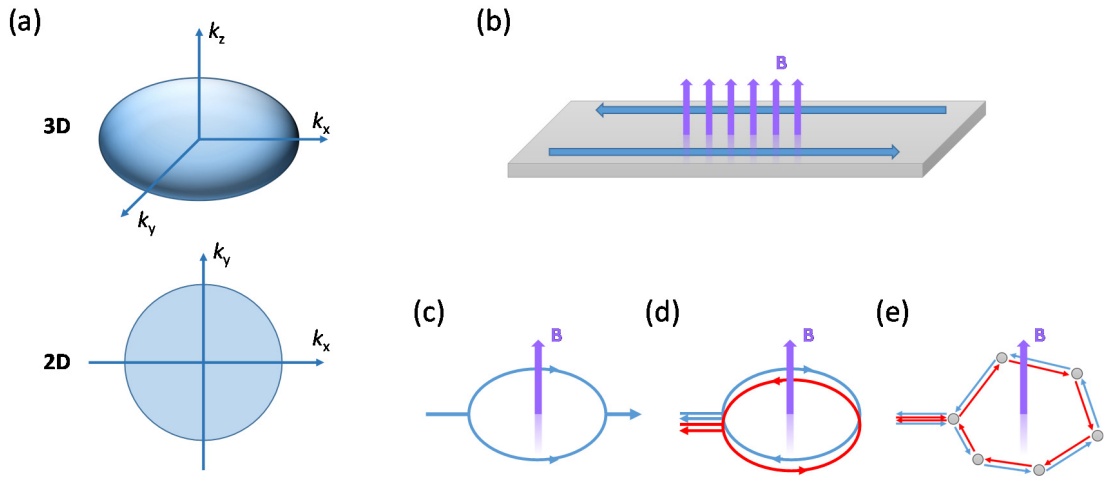

**Figure 2 (Color online). Transport methods for detecting the bulk and surface states.** (**a**) A comparison between 3D and 2D Fermi surfaces. They can be distinguished by mapping out the angle dependence of the cross-section area based on the SdH oscillations. (**b**) Illustration of the formation of the QH edge state under a high magnetic field. (**c-e**) Sketch of AB oscillations (**c**), AAS oscillations (**d**), and WL/WAL (**e**) in the presence of a magnetic field. The former occurs in the ballistic or pseudodiffusive transport regime, while the latter two occur in the diffusive transport regime. The difference between **d** and **e** is the sample geometry.

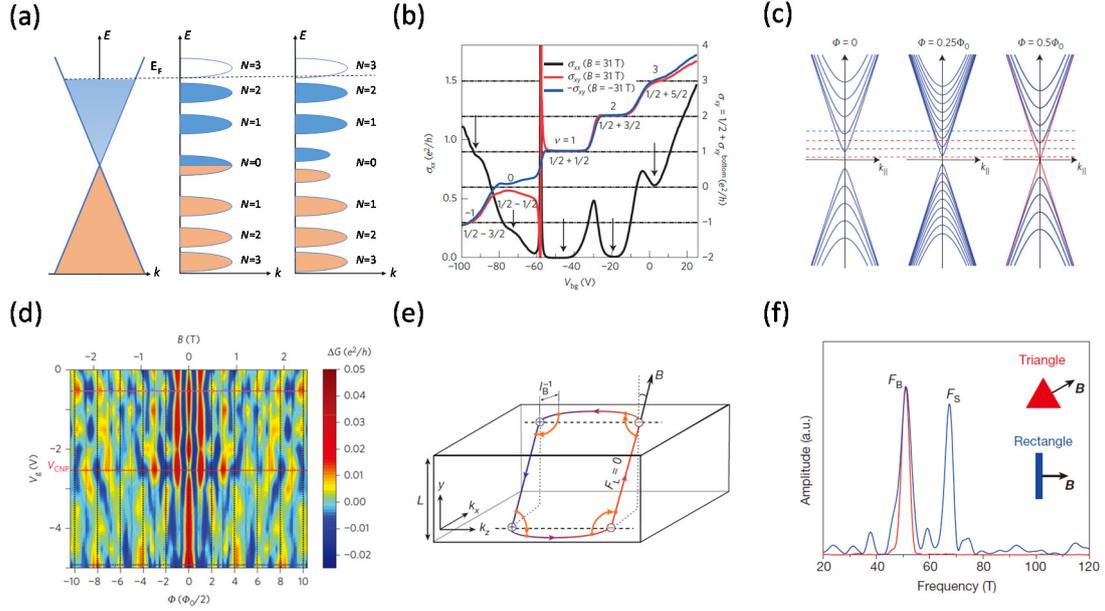

**Figure 3 (Color online). Detecting the band crossing point and Berry phase by transport.** (**a**) The Landau quantization of the Dirac cone (left) under a magnetic field. In a massless Dirac system (middle), the zeroth Landau levels of massless Dirac fermions remain at zero energy independent of the magnetic field. In massive Dirac systems (right), the zeroth Landau levels of conduction (marked in blue) and valence bands (marked in yellow) are separated by the energy gap. (**b**) Extracted 2D longitudinal and Hall conductivities ($\sigma_{xx}$ and $\sigma_{xy}$ at 31 T; $-\sigma_{xy}$ at -31 T). The plateaus observed in $\sigma_{xy}$ are labeled with the corresponding total Landau filling factors (the sum of the top and bottom surfaces). (**c**) The surface-state modes (subbands) of a TI nanowire for different magnetic fluxes along the wire. The blue lines are doubly degenerate, corresponding to clockwise and anti-clockwise circulation around the wire. The linear red curves are non-degenerate. When shifting the Fermi level, as indicated by the red and blue dashed lines, the total phase of the AB oscillations (the sum of the magnetic flux and the Berry phase of the surface modes) also changes, even if the magnetic flux is fixed. (**d**) Color plot of the AB oscillation amplitude $\Delta G$ versus the gate voltage $V_g$ and magnetic field flux at 0.25 K. The vertical dashed lines denote integer/half-integer flux quanta. (**e**) Sketch of the Weyl orbit in a thin slab of Weyl semimetal with a thickness of L under a magnetic field. **B**. The Weyl orbit consists of Fermi arcs on opposite surfaces and bulk chiral modes. The Weyl orbit includes the cyclotron orbit in momentum space and the propagating processes in real space. (**f**) Frequency spectrum of the triangular and rectangular samples, for field orientations perpendicular to each of the surfaces as shown in the insets. The surface oscillations disappear in the triangle samples because of the destructive interference of the quantum phase originating from the real-space propagating processes. Panel **b** is adapted from ref. [68], and panels **c-d** are adapted from ref. [97], panels **e-f** are adapted from ref. [64], Nature Publishing Group.

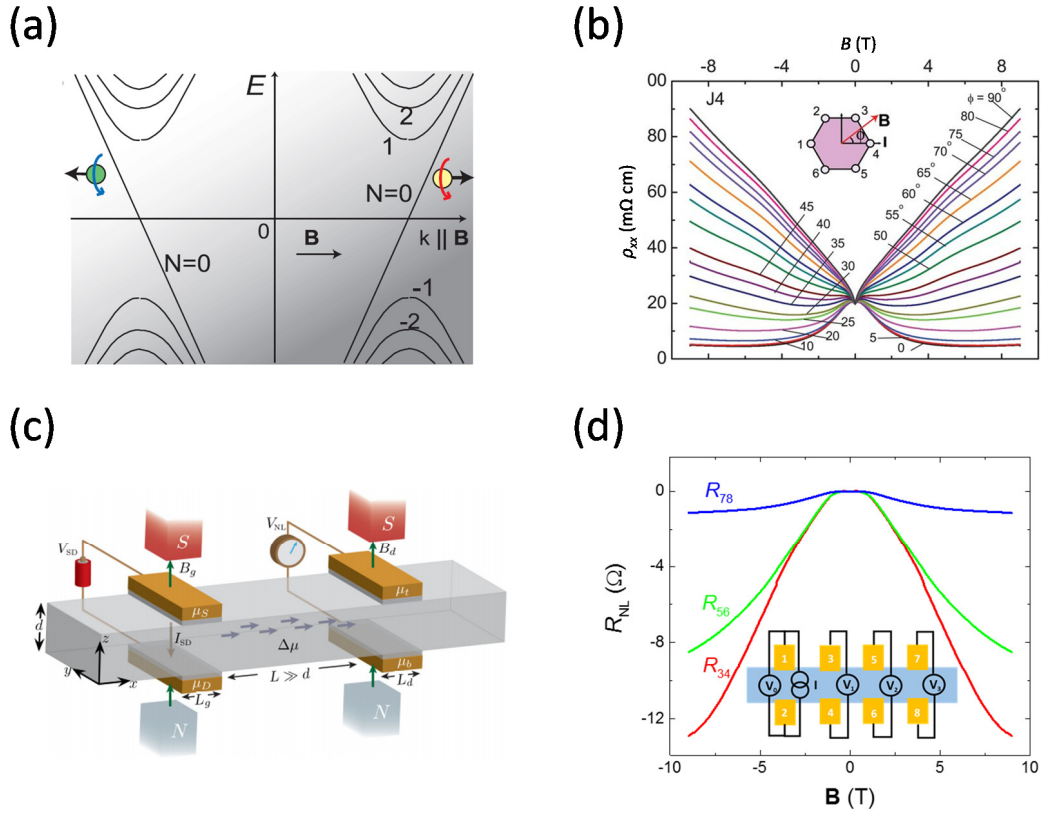

**Figure 4 (Color online). Chiral anomaly in TSs.** (**a**) A schematic of the chiral anomaly in Weyl semimetals. Parallel electric and magnetic fields cause an imbalance in the occupied density of states in the left-hand and right-hand Weyl cones. This imbalance breaks the conservation of the electron population for a fixed chirality and leads to a net chiral current anti-parallel to the electric field. (**b**) Magnetoresistivity curves of the Dirac semimetal Na$_3$Bi, with the magnetic field tilted from parallel (0°) to perpendicular (90°) relative to the applied current. A striking negative magnetoresistivity emerges at 0°. The inset shows the measurement geometry. (**c**) Proposed nonlocal transport experiment for detecting the chiral charge diffusion in TSs. (**d**) The nonlocal resistance $R_{NL}$ ($R_{34}$, $R_{56}$, and $R_{78}$) of Cd$_3$As$_2$ nanoplates at 100 K, which shows a magnetic field dependence opposite to that of the local resistance. The nonlocal resistance decays exponentially with increasing lateral length. The inset shows the device geometry. Panels **a-b** are adapted from ref. [115], AAAS. Panel **c** is adapted from ref. [124], American Physical Society. Panel **d** is adapted from ref. [119], Nature Publishing Group.

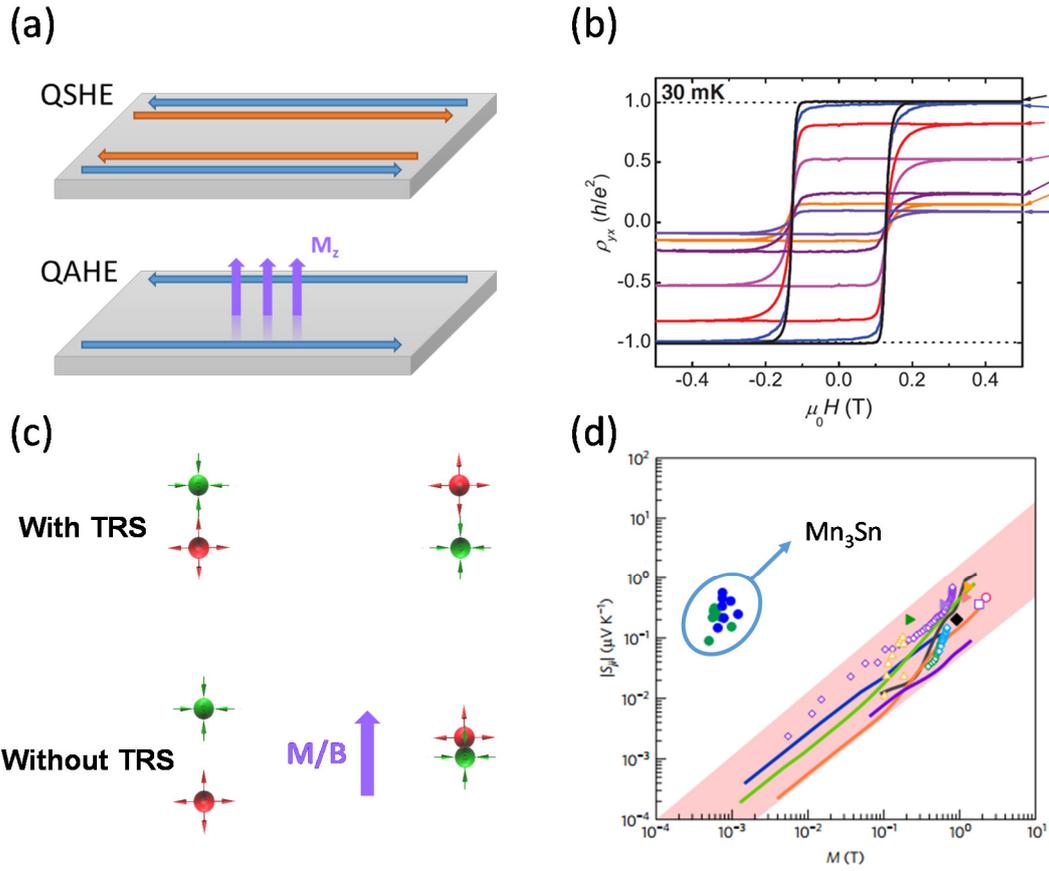

**Figure 5 (Color online). Breaking TRS in TIs and TSs.** (**a**) Illustration of the helical and chiral edge states in the QSHE and QAHE (or QHE if the magnetization is replaced with a magnetic field). The intrinsic magnetization is indicated by the purple arrows. The blue and orange lines correspond to spin-up and spin-down states, respectively. (**b**) The anomalous Hall effect $\rho_{yx}$ in Cr-doped $(Bi_{1-x}Sb_x)_2Te_3$ thin films. When the Fermi level is tuned inside the bulk gap, $\rho_{yx}$ becomes quantized. (**c**) Sketch of the shifting of Weyl node positions via magnetic coupling or the Zeeman effect. The sum of the wave vector between Weyl node pairs is zero in TSs with TRS and becomes nonzero when TRS is broken. (**d**) Magnetization dependence of the spontaneous Nernst effect for ferromagnetic metals and $Mn_3Sn$. Panel **b** is adapted from ref. [26], AAAS. Panel **d** is adapted from ref. [157], Nature Publishing Group.

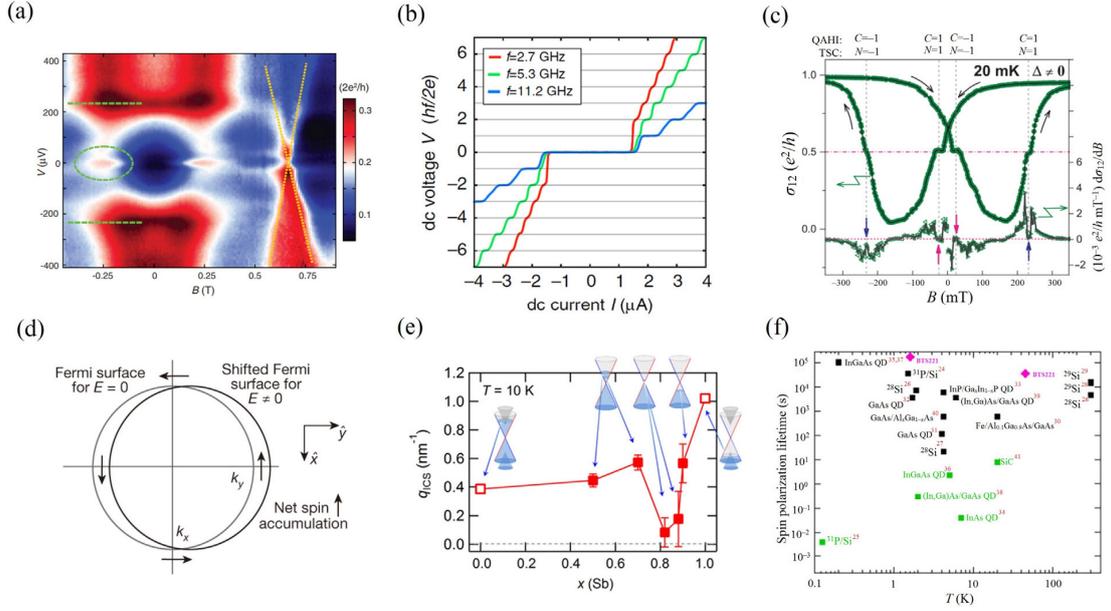

**Figure 6 (Color online). Manipulating topological systems for device applications.** (**a**) Color-scale plot of differential conductance versus $V$ and $B$. The ZBCP is indicated by the dashed oval. The green dashed lines denote the gap edges. (**b**) Shapiro steps of three different frequencies in a HgTe-based topological Josephson junction measured at $T$=800 mK. The $n$=1 step was missing at low frequency, while all other steps remained visible, implying the $4\pi$-periodic Josephson effect. (**c**). The half-integer QH plateau in the QAH-superconductor heterostructure. (**d**) Illustration of the generation of net spin polarization by a current-induced momentum shift in TI surface states. The arrows denote the directions of spin magnetic moments. (**e**) Interface charge-to-spin conversion efficiency for various Sb compositions in the $(Bi_xSb_{1-x})_2Te_3$/Cu/Py tri-layer heterostructures. The insets show the corresponding band structures and Fermi level positions for samples with different Sb ratios. (**f**) Comparison of the spin polarization lifetime of TIs with that of other systems. Panel **a** is adapted from ref. [163], panel **c** is adapted from ref. [179], panel **f** is adapted from ref. [188], AAAS. panel **b** is adapted from ref. [177], panel **d** is adapted from ref. [28], panel **e** is adapted from ref. [185], Nature Publishing Group.